\documentclass[twocolumn,showpacs,preprintnumbers,amsmath,amssymb]{revtex4}
\usepackage{graphicx}
\usepackage{dcolumn}
\usepackage{amsfonts}
\usepackage{amssymb}
\usepackage{epsfig}
\input epsf
\newcommand{\mbb}{\mathbb}
\newcommand{\ti}{\times}
\newcommand{\half}{\frac{1}{2}}
\newcommand{\mc}{\mathcal}

\newcommand{\fn}{\footnote}
\newcommand{\beqa}{\begin{eqnarray}}
\newcommand{\eeqa}{\end{eqnarray}}

\def\be{\begin{equation}}
\def\ee{\end{equation}}
\def\bea{\begin{eqnarray}}
\def\eea{\end{eqnarray}}

\begin{document}
\preprint{DAMTP-2006-110}
\preprint{hep-ph/0611144}
\title {\Large\bf The Neutrino Suppression Scale from Large Volumes }
\author{Joseph P. Conlon}
\author{Daniel Cremades}
\affiliation{%
DAMTP, Centre for Mathematical Sciences, Wilberforce Road, Cambridge
CB3 0WA, UK}%
\date{\today}
\begin{abstract}

We present an argument in which the scale $\sim 0.1$ eV associated
with neutrino masses naturally appears in a a class of
(very) large
volume compactifications, being tied to a supersymmetry scale of
$10^{3}$ GeV and a string scale of $10^{11} \hbox{GeV}$. The masses are of Majorana type and there is no
right-handed neutrino within the low-energy field theory.
The suppression scale $10^{14} \hbox{GeV}$ is independent of the masses of the heavy
states that are integrated out.
These kind of constructions appear naturally
in Type IIB flux compactifications. However, the arguments
that lead to this result rely only on a few geometrical features of
the compactification manifold, and
hence can be used independently of string theory.
\end{abstract}
\pacs{11.25.Mj 11.25.Wx 14.60.Pq 14.60.St}
\maketitle

\section{Introduction}
The existence of non-zero neutrino masses established in recent
years constitutes the
first evidence of physics beyond the Standard Model (for a review, see
\cite{hepph0510213}).
Neutrino mass splittings have been measured \cite{pdg}
\bea
7.7 \ti 10^{-5} ~\hbox{eV}^2 \lesssim & \Delta m_{12}^2 & \lesssim 8.4 \ti 10^{-5}~ \hbox{eV}^2, \nonumber \\
1.9 \ti 10^{-3} ~\hbox{eV}^2 \lesssim & \Delta m_{23}^2 & \lesssim 3.0 \ti 10^{-3}~ \hbox{eV}^2. \nonumber
\eea
The cosmological constraint $\sum_{\nu} m_{\nu}
\lesssim 0.7 \hbox{eV}$ thus implies the heaviest neutrino has a mass
\be
0.05~ \hbox{eV} \lesssim m_{\nu} \lesssim 0.3~ \hbox{eV}.
\ee
If this mass is of Dirac origin, the largest possible neutrino Yukawa coupling is
at least six orders of magnitude smaller than
that of the electron. This may be achieved in warped compactifications
\cite{hepph9806223, hepph9912408, hepph0312392}. However arguably the smallness of neutrino masses points to a
Majorana nature, as Majorana neutrinos require the existence of highly suppressed non-renormalisable operators.
This also implies the
existence of a new physical scale between the electroweak and Planck scales.

In the Standard Model, neutrino masses are generated after electroweak symmetry breaking by the
dimension five operator,
\be
\mc{O}_{5} = \frac{\lambda}{\Lambda} \left( L^T C i \tau_2 \tau L \right)
\left( H^T i \tau_2 \tau H \right).
\ee
In the MSSM, the comparable superpotential operator is (suppressing
gauge theory indices)
\be
\label{MSSMO}
\mc{O} = \frac{\lambda}{\Lambda} H_2 H_2 L L.
\ee
These are also the effective operators that appear in
the seesaw mechanism once the right-handed neutrinos are
integrated out. In the seesaw mechansim $\Lambda$ is the
scale of right-handed neutrino masses where new physics occurs.

The observed scale of neutrino masses creates a puzzle.
The MSSM Higgs vev is $\langle H_2 \rangle = v \sin \beta / \sqrt{2}$,
with $v = \hbox{246 GeV}$, which generates
neutrino masses
\be
m_{\nu} = \frac{\lambda v^2 \sin^2 \beta }{2 \Lambda} =  \lambda \sin^2 \beta
\left( 0.10~ \hbox{eV} \right) \left( \frac{3 \ti 10^{14} \hbox{GeV}}{\Lambda} \right)
\ee
If $\sin^2 \beta \sim 1$ (as applicable in the large $\tan
\beta$ regime),
$m_{\nu} \sim 0.1 \hbox{eV}$ requires a
suppression scale $\Lambda = 3 \ti 10^{14} \hbox{GeV}$ for $\lambda = 1$
(roughly the top Yukawa coupling) and $\Lambda \sim 3 \ti 10^{12} \hbox{GeV}$ for
$\lambda = 0.01$ (roughly the tau Yukawa coupling).

The value of $\Lambda$ is mysterious as it is neither the GUT
scale ($\sim 2 \ti 10^{16} \hbox{GeV}$), for which neutrino masses are
too small, nor the intermediate scale $10^{11} \hbox{GeV}$, for which
neutrino masses are too large: the scale $10^{12}
\hbox{GeV}  \div
10^{14} \hbox{GeV}$ does not correspond to any other
physical scale.

In this note we show that in a certain class of higher dimensional constructions,
and assuming only that lepton number is not a fundamental
symmetry of the low-energy theory, the required
scale of $\Lambda$
emerges naturally from requiring TeV-scale supersymmetry.
This class of
constructions appears naturally in the context of flux compactifications of Type IIB string
theory \cite{hepth0502058}, but the arguments leading to the correct scale of neutrino masses
need not be particular to them and rely only on a few geometric
features of the model.
Thus any construction, stringy or not, that shares these
characteristics will give this scale.

These constructions are characterised by a stabilised large compactification
manifold, with $\mc{V} \sim 10^{15}$ in string units, with an intermediate
string scale $m_s \sim 10^{11} \hbox{GeV}$. In this respect they have
similarities to the large extra dimensions scenario
\cite{hepph9803315, hepph9804398} (see \cite{hepph9811428, hepph9811448} for
neutrino masses in this scenario and \cite{hepph9809582,
  hepph9810535} for studies of an intermediate string scale). The Standard Model
is localised on branes wrapping small cycles within the large overall volume.
Considered in string theory, these constructions
present many appealing phenomenological features \cite{hepth0502058,
  hepth0505076}, such
as a naturally occurring hierarchy and computable flavour-universal
soft terms \cite{hepth0609180, hepth0610129}. The large volume also
allows a considerable degree of control and simplification in the computations.

The structure of this note is as follows.
In section II we review large volume models as presented in \cite{hepth0502058},
and estimate the strength of the $HHLL$ operator using the results of
\cite{hepth0609180}. In section III we show that the correct dimensionful suppression scale arises independently of the masses
of the heavy fields that are integrated out.
In section
IV we comment on embedding into a fully fledged string model.
In section V we conclude.

\section{Large Volume Models and Neutrino Masses}
\label{seLV}

The large volume models arise in IIB flux compactifications \cite{hepth0105097} with
$\alpha'$ corrections \cite{hepth0204254} and nonperturbative superpotential terms \cite{hepth0301240}.
The relevant K\"ahler potential and superpotential are \fn{${\cal V}$ is the dimensionless
volume, ${\cal V}={\rm Vol}/l_s^6$, with $l_s=2\pi\sqrt{\alpha'}$. $\Omega$ is the (3,0)-form of the Calabi-Yau $\mc{M}$, $S$ and $T$ are
dilaton and K\"ahler moduli superfields, $g_s$ is the string coupling and $G_3$ the 3-form flux.
$A_i$ and $a_i$ are $\mc{O}(1)$ numbers and $\xi \sim \chi(M)$
accounts for $\alpha'$ corrections.}
\bea
\mc{K} & = & - 2 \ln \left(\mc{V} + \frac{\xi}{g_s^{3/2}}\right) - \ln \left(i
\int \Omega \wedge \bar{\Omega} \right) - \ln (S + \bar{S}), \nonumber \\
W & = & \int G_3 \wedge \Omega + \sum A_i \exp (-a_i T_i). \label{super}
\eea
For the simplest model (the Calabi-Yau $\mbb{P}^4_{[1,1,1,6,9]}$), there are two K\"ahler moduli and the volume can be
written as
\be
\mc{V} = \frac{1}{9\sqrt{2}} \left( \left(\frac{T_b+\bar{T}_b}{2}\right)^{3/2} - \left(\frac{T_s+\bar{T}_s}{2}\right)^{3/2} \right).
\ee
Moduli stabilisation in this geometry has been studied in detail in \cite{hepth0502058, hepth0505076}.
The dilaton and complex structure moduli are flux-stabilised.
Denoting $\tau_s = \hbox{Re}(T_s), \tau_b = \hbox{Re}(T_b)$, the
scalar potential for the K\"ahler moduli is
\be
\label{pot}
V = \frac{\sqrt{\tau_s} a_s \vert A_s \vert^2 e^{-2 a_s
    \tau_s}}{\mc{V}} - \frac{a_s \tau_s \vert A_s W_0 \vert e^{-a_s \tau_s}}{\mc{V}^2}
+ \frac{\xi \vert W_0 \vert^2}{g_s^{3/2} \mc{V}^3}.
\ee
In this expression,  $W_0 = \langle \int G_3 \wedge \Omega \rangle$
and typically is $\mc{O}(1)$.
The minimum of (\ref{pot}) lies at exponentially large volume,
\be
\mc{V} \sim W_0 e^{c/g_s},
\ee
with $\tau_s \sim \ln \mc{V}$ and $c$ an $\mc{O}(1)$ constant. The
Standard Model lives on D7 branes wrapping the small cycle $\tau_s$.

The gravitino mass is given by
\be
\label{m32}
m_{3/2} = e^{K/2} W_0 = \frac{W_0}{\mc{V}}.
\ee
A TeV-scale gravitino mass, required for a natural solution to the weak hierarchy problem, requires a volume
$\mc{V} \sim 10^{15}$. This occurs for appropriate values of $g_s$ and $\xi$.

The extra-dimensional
geometry is then a small four-dimensional region containing the Standard Model, embedded in a large
six dimensional Calabi-Yau. We emphasize that while a natural stringy embedding exists, our results
will apply to any model giving this geometry.

The value of $\Lambda$ in the field theory equation (\ref{MSSMO})
depends on the underlying supergravity structure.
If we use $C$ to denote generic matter fields (and their scalar components),
on general grounds we can expand the superpotential as a power series in $C$,
\be
\label{ten}
W = \hat{W} + \frac{Y_{\alpha \beta \gamma}}{6} C^{\alpha} C^{\beta}
C^{\gamma} + \frac{Z_{\alpha \beta \gamma \delta}}{24 M_P} C^{\alpha} C^{\beta}
C^{\gamma} C^{\delta} + \ldots.
\ee
The dimensionful suppression scale in the superpotential must be $M_P$, as this is
the only scale known in supergravity. In stringy constructions this is directly required by
holomorphy: other scales such as the string or Kaluza-Klein scales are related to $M_P$ by
non-holomorphic functions of the moduli: e.g. $M_{s} \sim M_P / \mc{V}^{1/2} \sim M_P / (T_b+ \bar{T}_b)^{2/3}$.

Fermion bilinears arise via the
Lagrangian term
\be \int d^4 x \, e^{\hat{K}/2} \partial_{\alpha}
\partial_{\beta} W \psi^{\alpha} \psi^{\beta}.
\ee
This generates terms
\be
\label{unInt}
e^{\hat{K}/2} Y_{\alpha \beta \gamma} C^{\alpha} \psi^{\beta}
\psi^{\gamma} + e^{\hat{K}/2} \frac{Z_{\alpha \beta \gamma \delta}}{2 M_P}
C^{\alpha} C^{\beta} \psi^{\gamma} \psi^{\delta} \in \mc{L}
\ee
The
operator $\mc{O}$ thus arises from the superpotential term
$$
\frac{\lambda}{M_P} H_2 H_2 L L \in W.
$$
The main assumption we make is that, consistent with the
symmetries of the R-parity MSSM, this operator exists in
the superpotential.

Equation (\ref{unInt}) does not itself
give the physical coupling strength as in general
the matter field metrics fail to be canonically
normalised. The full Lagrangian is
\bea
\label{fullL}
\mc{L} & = &
\tilde{K}_{\alpha \bar{\beta}} \partial_\mu C^{\alpha} \partial^\mu \bar{C}^{\bar{\beta}} + e^{\hat{K}/2}
Y_{\alpha \beta \gamma} C^{\alpha} \psi^{\beta} \psi^{\gamma} +
 \nonumber \\
& & + e^{\hat{K}/2} \frac{Z_{\alpha \beta \gamma \delta}}{M_P} C^{\alpha} C^{\beta}
\psi^{\gamma} \psi^{\delta}.
\eea

The physical strengths of the dimension-four (i.e. Yukawa) and
dimension five couplings are then
\bea
\label{dim4}
\hat{Y}_{\alpha
\beta \gamma} & = & e^{\hat{K}/2} \frac{Y_{\alpha \beta
\gamma}}{(\tilde{K}_{\alpha} \tilde{K}_{\beta}
\tilde{K}_{\gamma})^{\half}}, \\
\label{dim5}
\hat{Z}_{\alpha \beta \gamma \delta} & = & e^{\hat{K}/2}
\frac{Z_{\alpha \beta \gamma \delta}}{M_P (\tilde{K}_{\alpha} \tilde{K}_{\beta}
\tilde{K}_{\gamma} \tilde{K}_{\delta})^{\half}}.
\eea
We have restricted (\ref{fullL}) to
  diagonal metrics, but this is only for convenience.
The K\"ahler potentials of
both moduli and matter fields are crucial for computing the physical
suppression scale of dimension-five operators.
In IIB string compactifications, the moduli
K\"ahler potential is
\be
\label{ModK}
\hat{K} = - 2 \ln \mc{V} - \ln
\left(i \int \Omega \wedge \bar{\Omega}\right) - \ln (S + \bar{S}).
\ee
The factor $-2 \ln \mc{V}$ is generic as it ensures
the `natural' scale of the scalar potential is $m_s^4$ rather than $M_P^4$ (using $m_s \sim \frac{M_P}{\sqrt{V}}$).
It will therefore be present in any model fitting into the framework described in section \ref{seLV}.

It was shown in \cite{hepth0609180} that for the models described in
section II,
the matter metric scales
at leading order as
\be
\label{MattK}
\tilde{K}_{\alpha \bar{\beta}} \sim \frac{\tau_s^{1/3}}{\mc{V}^{2/3}}
k_{\alpha \bar{\beta}}(\phi),
\ee
where $\phi$ are complex structure moduli.
The volume scaling is easiest to understand.
As the Standard Model interactions are localised
on a small cycle, the physical Yukawas must be
insensitive to the bulk volume. Using (\ref{dim4}) and (\ref{ModK}), this can hold
only if the matter metrics scale as $\tilde{K}_{\alpha} \sim \mc{V}^{-2/3}$ \fn{This assumes the superpotential Yukawas are independent of ${\cal V}$, which holds
for all stringy embeddings \cite{hepth0609180}.}. This is an argument from
locality, and so is not specific to string theory and will hold for any model giving this geometry.
The power of $\tau_s$ depends on the brane configuration -
1/3 is for the simplest case where all branes wrap the same cycle.

Equation (\ref{MattK}) suffices to determine the suppression scale for dimension-five operators. From (\ref{dim5}) we find
\be
\label{eqS}
\hat{Z}_{\alpha \beta \gamma \delta} \sim \frac{\mc{V}^{1/3}}{M_P \tau_s^{2/3}} Z_{\alpha \beta \gamma \delta}.
\ee
As in equation (\ref{m32}), we require $\mc{V} \sim 10^{15}$ for a natural solution to the weak hierarchy problem.
The size of the small cycle $\tau_s$
determines the Standard Model
gauge couplings by $\alpha_{SM}(m_s) \sim \tau_s^{-1}$ and so we require $\tau_s \sim 20$. The
dimension-five operator $\hat{Z}$ is then
\bea
\hat{Z}_{\alpha \beta \gamma \delta} & = &  \frac{Z_{\alpha \beta \gamma \delta}}{1.7 \ti 10^{14} \hbox{GeV}}  \left(
\frac{\mc{V}}{10^{15}} \right)^{1/3} \left(\frac{20}{\tau_s}\right)^{2/3}. \nonumber \\
\label{SuppScale}
\Lambda & = & \left( 1.7 \ti 10^{14} \hbox{GeV} \right) \left( \frac{10^{15}}{\mc{V}} \right)^{1/3} \left( \frac{\tau_s}{20} \right)^{2/3}
\eea
The operator $\hat{Z}_{\alpha \beta \gamma \delta}$ is defined at the string scale ($10^{11} \hbox{GeV}$) and
renormalised to the neutrino mass scale ($10^{-1}\rm{eV}$). MSSM loop corrections
involve weakly coupled gauge groups, and so while may alter the detailed form of $\hat{Z}_{\alpha \beta \gamma \delta}$ may alter, the
overall scale of $\Lambda$ should not change. This is analagous to the running
of slepton or sneutrino masses, which also get masses from non-renormalisable operators:
the physical and high-scale values differ by $\mc{O}(1)$ factors. Such a factor does not affect the overall
scale of (\ref{SuppScale}).
 If there exist extra strongly coupled hidden sectors,
 (\ref{SuppScale}) may be affected significantly: we assume this does not hold.

Equation (\ref{SuppScale}) is the main result of the paper.
For the models of section II, simply requiring a TeV gravitino mass, achieved by
$\mc{V} \sim 10^{15}$, naturally generates the required suppression scale $\Lambda \sim 10^{14} \hbox{GeV}$ for
neutrino masses. This result is appealing, as it ties the scale of neutrino masses to that of supersymmetry
breaking in a consistent manner.
It is also appealing that the scale of supersymmetry breaking is naturally hiearchically small,
as the moduli stabilisation mechanism of \cite{hepth0502058} generates
exponentially large volumes.

We can condense the arguments of this section into a
single formula relating neutrino masses, Higgs parameters, supersymmetry breaking
and the gauge coupling constants. Up to $\mc{O}(1)$ factors
in $Z_{\alpha \beta \gamma \delta}$, we find
\beqa
m_\nu \simeq \frac{v^2 \sin^2 \beta \left(
\alpha_{SM}(m_s) \right)^{2/3}}{2 M_P^{2/3}(m_{3/2})^{1/3}} .
\eeqa

\section{On Integrating Out Heavy Fields}

A surprising result is that although the string scale is $m_s \sim 10^{11} \hbox{GeV}$, the suppression scale is
$\Lambda \sim 10^{14} \hbox{GeV}$.
With an intermediate string scale $10^{11} \hbox{GeV}$ there exist heavy states integrated out in producing the
low energy theory. These include excited string states ($m_s \sim 10^{11} \hbox{GeV}$) and Kaluza-Klein states
($m_{KK} \sim 10^9 \hbox{GeV}$). Naively such states generate
dimension-five operators suppressed, in contrast to (\ref{SuppScale}), only by $m_s$ or $m_{KK}$.
However in fact the suppression scale (\ref{SuppScale})
arises independently of the mass scale that has been integrated out.

To show this, we imagine including
in the Lagrangian an explicitly massive field $\Phi$ with mass $m_{\Phi} \sim \mc{V}^{-\alpha} M_P$.
For stringy states $\alpha = 1/2$ and for KK states $\alpha = 2/3$.
The superpotential and K\"ahler potential are
\bea
\label{eqnn}
W & = & H_2 L \Phi + M_P \Phi^2, \\
K & = & \frac{\tau_s^{1/3}}{\mc{V}^{2/3}} H_2 \bar{H}_2 +  \frac{\tau_s^{1/3}}{\mc{V}^{2/3}} L \bar{L} + \mc{K}_{\Phi \bar{\Phi}} \Phi \bar{\Phi}.
\eea
Then $m_{\Phi}^2 = e^{\hat{K}} \frac{M_P^2}{\mc{K}_{\Phi \bar{\Phi}}^2}$ and
using (\ref{ModK}) $\mc{K}_{\Phi \bar{\Phi}} \sim \mc{V}^{\alpha - 1}$.
Neither $m_s$, $m_{KK}$ nor any function of $\mc{V}$ may
 enter (\ref{eqnn}) - as with (\ref{ten}), such a scale
would introduce a non-holomorphic dependence on the moduli.

The canonically normalised $H_2 L \Phi$ operator is
$$
\hat{Y}_{H_2 L \Phi} = e^{\hat{K}/2} \frac{Y_{H_2 L \Phi}}{\sqrt{K_{\Phi \bar{\Phi}} K_{L \bar{L}} K_{H_2 \bar{H}_2}}}
= \mc{V}^{1/6 - \alpha/2} Y_{H_2 L \Phi}.
$$

The $H_2 L H_2 L$ operator arises through integrating out $\Phi$ to generate a fermion-fermion-scalar-scalar
interaction, with an intermediate fermionic $\tilde{\Phi}$. With all fields canonically normalised, the fermionic propagator
is $\sim 1/m_{\Phi}$ and the resulting dimensionful interaction strength is $\mc{V}^{1/6 - \alpha/2} \ti \frac{\mc{V}^{\alpha}}{M_P} \ti
\mc{V}^{1/6 - \alpha/2} = \frac{\mc{V}^{1/3}}{M_P}$, independent of the mass scale integrated out.
This is exactly the suppression scale found in (\ref{eqS}). There is a sense
in which stringy or KK states play the role of right-handed neutrinos, although by their nature
such stringy fields cannot form part of the low-energy theory.

This result is striking independent of its relevance for neutrino masses:
the suppression scale of a dimension-five operator can be
entirely independent of the masses of the heavy states integrated out to produce the operator.

The above arguments apply to explicitly massive states such as stringy, Kaluza-Klein or winding states.
There may also exist dangerous heavy states with masses attained
in the field theory (e.g. through a Higgs mechanism): an example of such a state is a field-theoretic right-handed neutrino with a mass
of $10^5 \hbox{GeV}$. The existence of such states is model-dependent and we simply assume these are absent.

\section{On String Realisations}

In most of the original (Type II) semi-realistic string Standard Model constructions, starting with \cite{IMR},
lepton number is perturbatively conserved,
forbidding neutrino Majorana mass terms.
However these can be generated by either
GUT-like mechanisms (e.g. \cite{Cvetic:2006by}) or
nonperturbative brane instanton terms like $e^{-T} \nu_R \nu_R$ \cite{nu1,nu2}.
$U(1)_L$ is anomalous; the field $T$
is charged and transforms as
$T \to T - 2iQ$ while the neutrinos transform as $\nu_R\to e^{i Q}\nu_R$,
making this term gauge invariant.
Such an approach cannot be relevant here: both
because the instanton amplitudes $e^{-T} \sim \frac{1}{\mc{V}}$ are
too small and because the string scale $m_s \sim 10^{11} \hbox{GeV}$
restricts the maximal field theory mass of the right-handed neutrino.

We have instead assumed the $H_2 H_2
L L$ operator to be allowed - i.e. lepton number is not a perturbative symmetry - and computed its suppression scale.
The original work of \cite{IMR} notwithstanding,
recent approaches show that
there is no principled objection to
SM-like constructions in string theory violating lepton number \cite{Gepner}.

Given the absence of a conventional field theory see-saw
in large volume compactifications,
we find the attractive accordance of scales
found here notable. Following our arguments, for any large volume stringy construction
(in principle but not necessarily as in section II) in which the $HHLL$ operator enters
the superpotential with an
$\mc{O}(1)$ coefficient, the simple requirement of
TeV scale supersymmetry guarantees the correct scale for
neutrino masses.
We find this strong connection between TeV supersymmetry and
neutrino masses far from trivial, and regard it as motivation for the development
of global string models in which all the details take place.

\section{Conclusions}

This note argues that the mysterious energy scale
$\Lambda \sim 10^{14}{\rm GeV}$ of Majorana
neutrino masses has a clean origin, associated with geometric
features of a large class of compactification
manifolds. We require
(i) perturbative lepton number violation;
(ii) a geometry consisting of a
large six dimensional compactification manifold of volume ${\cal
  V}\sim 10^{15}$ in string units
(as required for
TeV-scale supersymmetry), with the
Standard Model
on a small four-dimensional
submanifold; (iii) a superpotential independent of $\mc{V}$,
as holds in string compactifications.

This suppression scale is independent of the masses of heavy states that are integrated out.
Given these assumptions one is led naturally to Majorana neutrino
masses of order $10^{-2}~{\rm eV} - 10^{-1}~{\rm eV}$.

\acknowledgments
We thank C. Burgess, L. Ib\'a\~nez and F. Quevedo for conversations. JC thanks the
Perimeter Institute and the ICTP, Trieste for hospitality while this
work was ongoing.

\end{document}